\documentclass[reprint,superscriptaddress,amsmath,amssymb,aps,pra,floatfix,longbibliography]{revtex4-2}
\usepackage{graphicx} 
\graphicspath{{figs/}} 
\usepackage{booktabs} 
\usepackage{bm} 
\usepackage{physics} 
\usepackage{hyperref} 
\usepackage{makecell} 

\hypersetup{%
     colorlinks   = true,
     breaklinks   = true,
     linkcolor    = {blue},
     anchorcolor  = {blue},
     filecolor    = {blue},
     citecolor    = {blue},
     urlcolor     = {blue},
}
\usepackage{xcolor}

\newcommand{\secref}[1]{\hyperref[#1]{Section~\ref{#1}}}
\newcommand{\appref}[1]{\hyperref[#1]{Appendix~\ref{#1}}}
\newcommand{\tabref}[1]{\hyperref[#1]{TABLE~\ref{#1}}}
\newcommand{\figref}[2][]{\hyperref[#2]{FIG.~\ref{#2}#1}}
\newcommand{\codeavailability}{\subsection*{CODE AVAILABILITY}}

\usepackage[normalem]{ulem} 
\usepackage{cancel} 

\usepackage{ifthen}
\newboolean{showchanges}
\setboolean{showchanges}{true} 

\DeclareRobustCommand{\add}[1]{%
    \ifthenelse{\boolean{showchanges}}%
    {{\color{blue}\uwave{#1}}}%
    {#1}%
}

\DeclareRobustCommand{\remove}[1]{%
    \ifthenelse{\boolean{showchanges}}%
    {{\color{red}\sout{#1}}}%
    {\relax}%
}

\DeclareRobustCommand{\blue}[1]{%
    \ifthenelse{\boolean{showchanges}}%
    {{\color{blue}#1}}%
    {#1}%
}

\DeclareRobustCommand{\red}[1]{%
    \ifthenelse{\boolean{showchanges}}%
    {{\color{red}#1}}%
    {#1}%
}

\begin{document}

%
\author{Pontus Vikst{\aa}l}
\email[e-mail:~]{vikstal@chalmers.se}
\affiliation{Wallenberg Centre for Quantum Technology, Department of Microtechnology and Nanoscience, Chalmers University of Technology, 412 96 Gothenburg, Sweden}

\author{Laura García-Álvarez}
\affiliation{Wallenberg Centre for Quantum Technology, Department of Microtechnology and Nanoscience, Chalmers University of Technology, 412 96 Gothenburg, Sweden}

\author{Shruti Puri}
\affiliation{Department of Applied Physics, Yale University, New Haven, Connecticut 06511, USA}
\affiliation{Yale Quantum Institute, Yale University, New Haven, Connecticut 06511, USA}

\author{Giulia Ferrini}
\affiliation{Wallenberg Centre for Quantum Technology, Department of Microtechnology and Nanoscience, Chalmers University of Technology, 412 96 Gothenburg, Sweden}

%
\title{Quantum Approximate Optimization Algorithm with Cat Qubits}

%
\date{\today}

\begin{abstract}
    The Quantum Approximate Optimization Algorithm (QAOA)---one of the leading algorithms for applications on intermediate-scale quantum processors---is designed to provide approximate solutions to combinatorial optimization problems with shallow quantum circuits. Here, we study QAOA implementations with cat qubits, using coherent states with opposite amplitudes. The dominant noise mechanism, i.e., photon losses, results in $Z$-biased noise with this encoding. We consider in particular an implementation with Kerr resonators. We numerically simulate solving MaxCut problems using QAOA with cat qubits by simulating the required gates sequence acting on the Kerr non-linear resonators, and compare to the case of standard qubits, encoded in ideal two-level systems, in the presence of single-photon loss. Our results show that running QAOA with cat qubits increases the approximation ratio for random instances of MaxCut with respect to qubits encoded into two-level systems.
\end{abstract}

%
\maketitle

\section{\label{sec:Introduction}Introduction}
Variational quantum algorithms~\cite{Cerezo2021,Montanaro2016}, combining quantum and classical computation in a hybrid approach, occupy a central role in current research on quantum algorithms. These algorithms are promising for implementations on NISQ devices~\cite{Preskill18}, since they can in principle run on shallow quantum processors. In particular, the Quantum Approximate Optimization Algorithm (QAOA)~\cite{farhi2014quantum} can be used to tackle combinatorial optimization problems, which are omnipresent in logistics, with applications within the automotive sector~\cite{streif2020beating,fitzek2021applying}, or aviation, e.g., aircraft~\cite{vikstal2020applying} or gate~\cite{stollenwerk2019flight} assignment, financial portfolio optimization~\cite{hodson2019portfolio}, among others. First proof-of-principle implementations of QAOA in superconducting qubit devices were used to solve MaxCut~\cite{otterbach_unsupervised_2017,arute2020quantum} and Exact Cover~\cite{bengtsson2020improved,lacroix2020improving} problems. Although the performance of QAOA improves at increasing algorithmic depth provided optimal parameters, current NISQ hardware is limited by noise, which decreases the performance of QAOA after a certain algorithmic depth~\cite{arute2020quantum}. As such, research into different avenues for hardware implementations of QAOA that could allow for reaching deeper circuits is needed. 

In this work, we explore the implementation of QAOA in bosonic systems.  These have led to promising quantum computing implementations in a variety of physical settings including optical~\cite{pfister2019continuous} and microwave radiation~\cite{grimsmo2017squeezing, hillmannUniversalGateSet2020,campagne-ibarcq_quantum_2020}, trapped ions~\cite{serafini2009manipulating, fluhmann2019encoding, de_neeve_error_2020}, opto-mechanical systems~\cite{schmidt2012optomechanical, houhou2015generation, nielsen2017multimode}, atomic ensembles~\cite{stasinska2009manipulating, milne2012composite, ikeda2013deterministic, motes2017encoding}, and hybrid systems~\cite{aolita2011gapped}. For example, in the microwave regime, bosonic codes have successfully extended the life-time of quantum information in superconducting cavities compared to the system's constituents~\cite{ofek2016,ni_beating_2022,sivak_real-time_2022}.

So far, bosonic implementations of QAOA have primarily focused on optimizing continuous functions defined on real numbers~\cite{verdon2019quantum, Yutaro}, with little attempt made to address QAOA for solving discrete optimization problems in the context of bosonic systems, which is the focus of our work.

Encoding qubits into the coherent states of cavities fields $\ket{\pm \alpha}$, yielding cat qubits, is an emerging approach that results in biased noise. Such type of noise affects a quantum system in a non-uniform way, i.e., certain types of errors are more likely to occur than others. This encoding can thus lead to favorable error-correcting properties~\cite{aliferis_fault_2008,tuckett_ultrahigh_2018}, and to enhanced algorithmic performance~\cite{Gouzien2023}.

In a previous work~\cite{vikstal_study_2022}, some of the authors have shown that biased-noise qubits also allow for implementing error mitigation techniques and achieving higher performance ratios in QAOA compared to standard qubits. However, those results were obtained for a generic noise-biased error model, without considering specific implementations.  In this work, we explore QAOA using cat qubits, achieved in particular by means of the driven Kerr non-linear resonator~\cite{puri_annealing_2017}. First, we simulate solving a two-qubit Exact Cover problem under the full master equation with cat qubits as a proof of principle demonstration. Second, to simulate larger cat qubit systems, we use the Pauli-transfer matrix formalism to characterize the error channel induced by single-photon losses on the computational subspace. We numerically show that for 6, 8 and 10-qubit MaxCut problems the use of cat qubits yields an improvement of the algorithmic approximation ratio with respect to the case of qubits encoded into discrete two-level systems, given equal average gate fidelities between the two systems. While we are going to focus on driven Kerr-nonlinear resonator, the implementation of QAOA on cat qubits yielding enhanced algorithmic performance unveiled in our work could also be achieved by means of other platforms, both in the superconducting~\cite{grimm_stabilization_2020}, as well as photonics~\cite{takase_generation_2022}, or other bosonic systems~\cite{cosacchi_schrodinger_2021}.

The paper is structured as follows. In \secref{sec:cat_qubits} we recall the definition of cat qubits as well as the gates needed to operate them. In \secref{sec:QAOA-Kerr} we outline how QAOA can be run on cat qubits. We first show the principle by considering a two-qubit toy model for solving the Exact Cover problem, and then consider more extensive simulations up to 8 qubits for solving MaxCut, in the presence of photon losses. We then compare the performance of QAOA with cat qubits to the one with standard qubits given the same average gate fidelity of the two systems. We provide our conclusive remarks in \secref{sec:Conclusion}. In Appendix \ref{app:avg-gate-fid} we recall the definition of quantum gates acting on cat qubits. In Appendix \ref{app:approxr} we provide some details regarding the numerical optimization. Finally, in Appendix \ref{app:bosonic QAOA} we introduce a bosonic version of QAOA by Trotterizing the relevant quantum annealing Hamiltonian, and we compare its performance to QAOA for the case of a single Ising spin.

\section{Cat qubits and how to operate on them}
\label{sec:cat_qubits}
In this section we recall the main properties of cat qubits implemented by means of the Kerr nonlinear resonator (KNR) as introduced in Refs.~\cite{goto_universal_2016, puri_engineering_2017}, and summarize how to perform gates on such a cat qubit, aiming at providing a self-consistent introduction.

\subsection{\label{sec:cat-qubit}The Kerr nonlinear resonator}
The cat qubit can be realized in a Kerr parametric oscillator with a two-photon pump~\cite{goto_universal_2016,grimm_stabilization_2020,puri_engineering_2017,puri_bias_2020}. In a frame rotating at the frequency of the two-photon pump and in the rotating-wave approximation, the Hamiltonian for a KNR is given by (we use $\hbar=1$ throughout this paper)
\begin{equation}
    \hat H_1 = - \Delta \hat a^\dagger \hat a - K \hat a^{\dagger 2} \hat a^2 + G(\hat a^{\dagger 2}e^{i2\phi}+\hat a^2e^{-i2\phi}),
\end{equation}
where $\Delta=\omega_r-2\omega_p$ is the detuning of the resonator frequency from twice the two-photon pump frequency, $K$ is the amplitude of the Kerr non-linearity, $G$ and $\phi$ are the amplitude and phase of the two-photon drive respectively. We assume that $K$ is a nonzero positive constant and that $\Delta$ is non-negative. When the detuning is zero (i.e. when the two-photon drive frequency is half the resonator frequency) and when the phase $\phi$ is zero, the KNR Hamiltonian can be written as
\begin{align}
\label{eq:KNR-Hamiltonian}
    \hat H_1 &= - K \hat a^{\dagger 2} \hat a^2 + G(\hat a^{\dagger 2}+\hat a^2) \nonumber \\
    &= - K \left(\hat a^{\dagger 2}-\frac{G}{K}\right) \left(\hat a^2-\frac{G}{K}\right)+\frac{G^2}{K}.
\end{align}
Since $\hat a\ket{\alpha}=\alpha\ket{\alpha}$, the coherent states $\ket{\pm\alpha}$ with $\alpha=\sqrt{G/K}$ are degenerate eigenstates of the Hamiltonian in Eq.~\eqref{eq:KNR-Hamiltonian} with eigenenergy $G^2/K$. The combinations of these degenerate eigenstates given by $\ket{C^\pm_\alpha}=N_\pm(\ket{\alpha}\pm\ket{-\alpha})$ with $N_\pm = \sqrt{2(1\pm e^{-2\abs*{\alpha^2}})}$ are the so-called cat states. These states are also degenerate eigenstates and have a well defined even and odd parity, with the parity operator $\hat \Pi=e^{i\pi\hat a^\dagger \hat a}$.

\begin{figure}[!htp]
    \centering
    \includegraphics[width=0.8\linewidth]{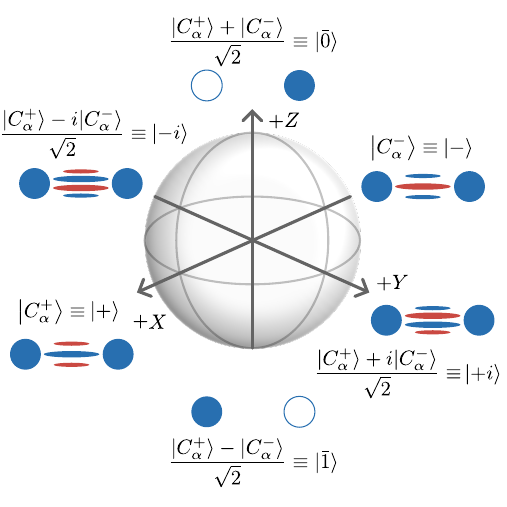}
    \caption{The computational states that lie along the $x,y,z$-axis implemented with cat qubits and visualized on the Bloch sphere along with their Wigner function.}
    \label{fig:Bloch}
\end{figure}

We can take advantage of this well-defined subspace to encode our computational basis states $\ket{\bar 0}$, $\ket{\bar 1}$, defining the qubit (the bar notation is used to distinguish the computational states from the zero and one photon Fock state). To this aim, one possibility is to directly identify the qubit basis states with $\ket{\alpha}$ and $\ket{-\alpha}$~\cite{puri_annealing_2017}. However, these states are quasi-orthogonal as $\braket{-\alpha}{\alpha}=e^{-2\alpha^2}$, and only become orthogonal in the high photon number limit. Another possibility consists in choosing the following encoding~\cite{puri_stabilized_2019}:
\begin{equation}
    \label{eq:kerrqubit}
    \ket{\bar 0} = \frac{\ket{C^+_\alpha}+\ket{C^-_\alpha}}{\sqrt{2}},
    \quad
    \ket{\bar 1} = \frac{\ket{C^+_\alpha}-\ket{C^-_\alpha}}{\sqrt{2}}.
\end{equation}
In this case, the computational basis states are orthogonal even for small $\alpha$, while for large $\alpha$ they are approximately equal to $\ket{\bar 0}\approx\ket{\alpha}$ and $\ket{\bar 1}\approx\ket{-\alpha}$. For single-photon losses, the encoding of Eq.~\eqref{eq:kerrqubit} constitutes a noise-biased qubit where the loss of a single-photon results in a phase error plus an exponentially small bit-flip error on the computational states with respect to $\alpha$. Indeed, by defining the projection operator $\hat I = \dyad{\bar 0} + \dyad{\bar 1}$, its action on the annihilation operator $\hat a$ gives
\begin{equation}
    \hat I \hat a \hat I = \frac{\alpha}{2}(\eta+\eta^{-1})\hat {Z} + i\frac{\alpha}{2}(\eta-\eta^{-1}) \hat{Y},
\end{equation}
where $\eta\equiv N_+ / N_-$, and $\hat{Z}$, $\hat{Y}$ are the two Pauli matrices in the computational subspace. For large $\alpha$, $\eta\rightarrow 1$, which results in $\hat I \hat a \hat I=\alpha \hat Z$, and we thus see that a single-photon loss event corresponds to a phase-error on the computational basis states. We will refer to the encoding in Eq.~\eqref{eq:kerrqubit} as the cat qubit. The computational basis states are shown on the Bloch sphere in \figref{fig:Bloch}.

In order to run QAOA, one needs to prepare all resonators in state $\ket{+}$, i.e., in the case of the cat qubit, the cat state $\ket{C^+_\alpha}$. Such a cat state can be generated deterministically in KNRs by starting from the vacuum, which is an eigenstate of Hamiltonian Eq.~\eqref{eq:KNR-Hamiltonian} for $G=0$, and then adiabatically increasing $G$~\cite{goto_bifurcation_2016,puri_engineering_2017}. Since the Hamiltonian in Eq.~\eqref{eq:KNR-Hamiltonian} is symmetric under parity inversion $\hat a \rightarrow -\hat a$, the KNR follows the adiabatic evolution from the vacuum while also conserving the parity, $[\hat \Pi,\hat H]=0$, thus ending up in the even parity cat state $\ket{C^+_\alpha}$. Alternatively, a cat state can also be generated using a sequence of SNAP and displacement gates applied to the vacuum state~\cite{marina_robust_2022}.

\subsection{Set of universal gates on the cat qubit}
\label{sec:gates-cat-qubit}
\begin{figure}[htbp]
    \centering
    \includegraphics[width=0.8\linewidth]{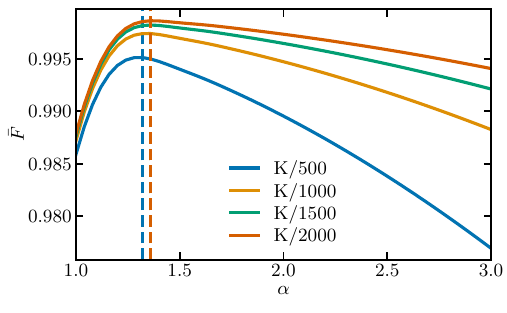}
    \caption{Average gate fidelity of the $R_Z(\phi)$-gate as a function of the coherent state amplitude $\alpha = \sqrt{G/K}$, for various single-photon loss rates. The optimal $\alpha$ that results in the maximum average gate fidelity for each loss rate is highlighted with a dashed vertical line. For loss rates varying from $K/1000$ to $K/2000$, the optimal $\alpha$ that achieves the highest average gate fidelity is $\alpha = 1.36$. Conversely, for a loss rate of $K/500$, the ideal $\alpha$ is $1.32$.}
    \label{fig:rz-alpha}
\end{figure}

We are now interested in the implementation of gates on the cat qubit. We are going to focus on the following gate set:
\begin{align}
    \label{eq:rz-gate}
    R_Z(\phi) &= e^{-i\phi \hat Z/2},
    \\
    \label{eq:rx-gate}
    R_X(\theta) &= e^{-i\theta \hat X/2},
    \\
    \label{eq:ry-gate}
    R_Y(\varphi) &= e^{-i\varphi \hat Y/2},
    \\
    \label{eq:rzz-gate}
    R_{ZZ}(\Theta) &= e^{-i\Theta \hat Z_1 \hat Z_2/2},
\end{align}
where $\{\hat X,\hat Y,\hat Z\}$ are the Pauli matrices in the computational basis, which in this case is taken to be the cat qubit Eq.~\eqref{eq:kerrqubit}. Note that this is an over-complete gate set, as any pair of single-qubit gates $\{R_X(\theta),R_Y(\varphi),R_Z(\phi)\}$ together with $R_{ZZ}(\Theta)$ allow for implementing arbitrary qubit operations. The gates are implemented according to Refs.~\cite{goto_universal_2016,puri_engineering_2017}, where the $R_Z(\phi)$-gate is implemented in KNRs by means of a single-photon drive. The $R_X(\theta)$-gate is implemented through a time-dependent detuning $\Delta$. The $R_Y(\varphi)$-gate is implemented by means of single and two-photon drives, and $R_{ZZ}(\Theta)$-gate is implemented through a beam-splitter interaction between two KNRs. We provide a more detailed description of these gates in \appref{app:avg-gate-fid}, where we also present numerical simulations validating this approach for relevant parameter regimes and in presence of noise induced by single-photon loss. In \mbox{\figref{fig:rz-alpha}}, we plot the average gate fidelity for the $R_Z(\phi)$-gate as a function of the coherent state amplitudes $\alpha$, and consider different loss rates to determine the value of $\alpha$ that yields the highest average gate fidelity. In \tabref{tab:avg-fidelity} we report the average gate fidelities, both without single-photon loss and with a single-photon loss rate of $K/1500$, specifically for the $\alpha$ that yields the optimal gate fidelity in that case, $\alpha = 1.36$. As a consequence of the noise bias and shorter gate-time, we observe that for the cat qubits, the two-qubit $R_{ZZ}$ gate exhibits a higher average gate fidelity compared to the single-qubit $R_X$ and $R_Y$ gates. However, when comparing the noise-bias preserving single-qubit $R_Z$ gate to the two-qubit $R_{ZZ}$ gate, the single-qubit $R_Z$ gate demonstrates a higher average gate fidelity, as expected.
\begin{table}[!htb]
    \centering
    \caption{Average gate fidelities for the considered gates within KNR-encoding obtained through master equation simulation. The results are averaged over 20 points evenly spaced between 0 and $\pi$. We consider a single-photon loss rate of $K/1500$ and the corresponding optimal coherent state amplitude $\alpha = 1.36$.}
    \begin{ruledtabular}
        \begin{tabular}{l c c}
        Gate & \makecell[c]{Avg. gate fid. (\%) \\ with no loss} & \makecell[c]{Avg. gate fid. (\%) \\ with single-photon loss} \\ \midrule
        $R_Z(\phi)$ & $>99.99$ & 99.64 \\ 
        $R_X(\theta)$ & $>99.99$ & 98.59 \\ 
        $R_Y(\varphi)$ & $\phantom{>}~99.52$ & 98.72 \\ 
        $R_{ZZ}(\Theta)$ & $>99.99$ & 99.15 \\ 
        \end{tabular}
        \end{ruledtabular}
    \label{tab:avg-fidelity}
\end{table}

\section{\label{sec:QAOA-Kerr}QAOA with cat qubits}
In this section, we use the gate set defined in \secref{sec:gates-cat-qubit} to implement the QAOA sequence on cat qubits. We start by briefly reviewing QAOA, and we then address numerical simulations of increasing complexity (two to eight qubits) in the presence of single-photon losses, assessing the algorithmic performance in terms of the success probability and the approximation ratio.
\subsection{\label{sec:qaoa}The QAOA algorithm}
QAOA~\cite{farhi2014quantum} starts from the superposition of all possible computational basis states, $\ket{+}^{\otimes n}$, where $n$ is the number of qubits. Then the alternating sequence of the two parametrized non-commuting quantum gates $\hat{U}(\gamma)$ and $\hat{V}(\beta)$ is applied $p$ times, with
\begin{align}
    \hat{U}(\gamma)\equiv e^{-i\gamma \hat{H}_C},\quad\hat{V}(\beta) \equiv e^{-i\beta \hat{H}_M},
\end{align}
where $\hat{H}_M\equiv\sum_{i=1}^n\hat X_i$ is the mixing Hamiltonian, and $\hat{H}_C$ is the cost Hamiltonian that encodes the solution to the considered optimization problem in its ground state,
\begin{equation}
    \label{eq:cost-hamiltonian}
    \hat H_C = \sum_{i<j} J_{ij}\hat Z_i\hat Z_j + \sum_i h_i\hat Z_i.
\end{equation}
Indicating the collection of variational parameters as $\vec{\gamma}=(\gamma_1,\ldots,\gamma_p)$ with $\gamma_i\in[0,2\pi)$ if $\hat{H}_C$ has integer-valued eigenvalues, and $\vec{\beta}=(\beta_1,\ldots,\beta_p)$ with $\beta_i\in[0,\pi)$, the final variational state is
\begin{equation}
    \label{eq:final-state}
    \ket*{\psi_p(\vec{\gamma},\vec{\beta})} \equiv \hat{V}(\beta_p)\hat{U}(\gamma_p)\ldots \hat{V}(\beta_1)\hat{U}(\gamma_1)\ket{+}^{\otimes n}.
\end{equation}
The parametrized quantum gates are then optimized in a closed loop using a classical optimizer with the objective of minimizing the expectation value of the cost Hamiltonian
\begin{equation}
    \label{eq:opt-parameters}
    (\vec{\gamma}^*,\vec{\beta}^*)=\mathrm{arg}\min_{\vec{\gamma},\vec{\beta}}\matrixelement*{\psi_p(\vec{\gamma},\vec{\beta})}{\hat{H}_C}{\psi_p(\vec{\gamma},\vec{\beta})}.
\end{equation}
Once the optimal variational parameters are found, one samples from the state $\ket*{\psi_p(\vec{\gamma}^*,\vec{\beta}^*)}$ by measuring it in the computational basis, the eigenvalue of the cost Hamiltonian Eq.~\eqref{eq:cost-hamiltonian} corresponding to the measured configuration, is evaluated. The success probability is defined as the probability of finding the qubits in the ground state configuration when performing a single shot measurement of the $\ket*{\psi_p(\vec{\gamma},\vec{\beta})}$ state, i.e.
\begin{equation}
    \label{eq:fidelity}
    F_p(\vec{\gamma},\vec{\beta}) \equiv \sum_{z_i\in\vec{z}_\mathrm{sol}}|\braket*{z_i}{\psi_p(\vec{\gamma},\vec{\beta})}|^2,
\end{equation}
where $z_i$ is a bit-string of length $n$, and $\vec{z}_\mathrm{sol}$ is the set of all bit string solutions.

It is clear that QAOA can be run on cat qubits and compiled using the gates discussed in \secref{sec:gates-cat-qubit}. The unitary $e^{-i\beta\hat H_M}$ can easily be implemented as single qubit $R_{X}(2\beta)$-gates on each individual qubit, and the cost Hamiltonian $\hat H_C$ can be implemented as a product of $R_{Z}(2\gamma h_i)$-gates and $R_{ZZ}(2\gamma J_{ij})$-gates~\cite{vikstal_application_2020}.
\subsection{\label{sec:qaoa-toy}Solving a toy problem with QAOA on cat qubits}
In order to test the capability of cat qubits for solving combinatorial optimization problems using QAOA given relevant gate fidelities for the set of operations considered, we run a master equation simulation of a two-qubit Exact Cover problem on cat qubits. 

Exact Cover is an NP-complete problem~\cite{Karp1972,Garey1979} that appears in logistics, and notably as a part of the Tail Assignment problem~\cite{vikstal2020applying}. The Exact Cover is formulated as follows: given a set $U=\{c_1,c_2,\ldots,c_n\}$, and a set of subsets $V=\{V_1,\ldots,V_m\}$ with $V_i\subset U$ such that
\begin{equation}
    U=\bigcup_{i=1}^m V_i,
\end{equation}
the goal is to decide if there exists a subset of the set of sets $\{V_i\}$, called $R$, such that the elements of $R$ are disjoint sets, i.e., $V_i\cap V_j=\emptyset$ for $i\neq j$, and the union of elements of $R$ is $U$.

\begin{figure}
    \centering
    \includegraphics{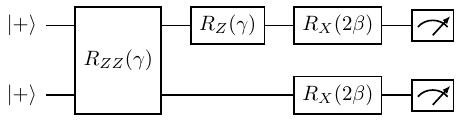}
    \caption{The circuit diagram of QAOA with depth $p=1$ for solving a two-qubit instance of the Exact Cover problem, using the universal gate set introduced in Eq.~\eqref{eq:rz-gate}-\eqref{eq:rzz-gate}. Here, the circuit is shown using the $\hat X$-mixer.}
    \label{fig:circuit-diagram}
\end{figure}

\begin{figure}[!t]
    \centering
    \includegraphics[width=0.8\linewidth]{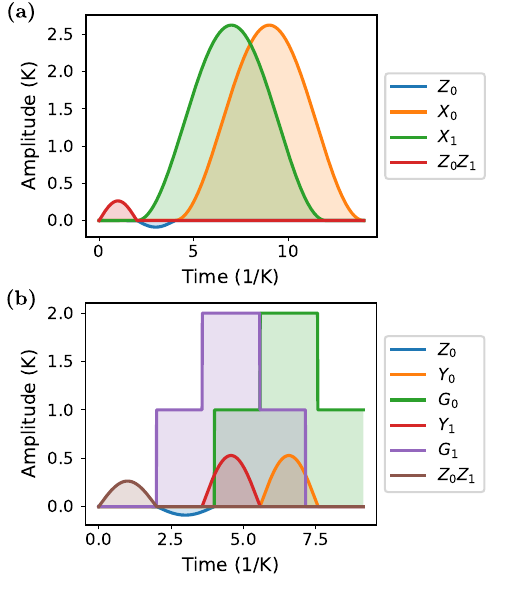}
    \caption{QAOA depth $p=1$ pulse schedule and shape \textbf{(a)} with $X$-mixer, \textbf{(b)} with $Y$-mixer. Each label corresponds to a Hamiltonian, for example $Z_0$ corresponds to the amplitude in units of the Kerr non-linearity of the Hamiltonian that implements the $R_Z$-gate on the zero-th cat qubit. Furthermore, the $G$-label in \textbf{(b)} corresponds to the amplitude of the two-photon drive, where the unit amplitude corresponds to a net two-photon drive of zero amplitude and the two unit amplitude corresponds to two-photon driving along the $P$-quadrature. This is because, in the simulations, the two-photon drive is always on. This is not shown in the figure, just as the always present self-Kerr. Therefore, to turn off the always present two-photon drive, an additional two-photon drive Hamiltonian is turned on, but with an opposite amplitude. A more detailed description of how the gates are implemented can be found in \appref{app:avg-gate-fid}.}
    \label{fig:pulse}
\end{figure}

For two qubits, the simulation of the circuit with the action of the gates can be carried out by solving the Lindblad master equation for the Kerr resonators. Therefore, we start by simulating Exact Cover for the same toy instance that was considered in Ref.~\cite{bengtsson2020improved}, i.e. $U=\{c_1,c_2\}$ and $V=\{V_1,V_2\}$, with $V_1=\{c_1,c_2\}$ and $V_2=\{c_2\}$. This has solution $\ket{\bar 1 \bar 0}$, corresponding to choosing subset $V_1$. The mapping onto the cost Hamiltonian Eq.~\eqref{eq:cost-hamiltonian} gives us the values $h_1=1/2$, $h_2=0$ and $J_{12}=1/2$~\cite{vikstal2020applying}. Therefore, the quantum circuit for implementing QAOA with $p=1$ takes the form of the one in \figref{fig:circuit-diagram}.
We extend our analysis of the original QAOA proposal and allow for different input states, namely $\ket{+}$ and $\ket{+i}$, and mixing Hamiltonians $\hat H_M$. Specifically we do simulations for both $\hat X$ and $\hat Y$-mixer, which corresponds to replacing the $R_X(\theta)$-gate with a $R_Y(\theta)$-gate in \figref{fig:circuit-diagram}. For simplicity, in the simulation contained in this self-contained  sub-section, dealing anyway with a small-size toy model, we chose a non-optimal coherent state amplitude of $\alpha=2$. \figref{fig:pulse} illustrates the amplitude of the pulse schedule for $p=1$ with the $\hat X$ and $\hat Y$ mixer respectively for the gates introduced in \secref{sec:gates-cat-qubit}. We simulate QAOA implemented with cat qubits using the numerically best found variational parameters $(\vec{\gamma},\vec{\beta})$ for the ideal, no losses case, up to $p=2$. The reason for using the variational parameters for the ideal case is that several results have shown that the optimal variational parameters are robust to noise~\cite{xue_effects_2019,sharma_noise_2020}, and because it is computational expensive to perform an extensive global optimization simulation of the system.

The results are summarized in \tabref{tab:results}. First of all, we observe that as a general result (independent of the cat qubit implementation), if the initial state is not an eigenstate of the mixer Hamiltonian, $100\%$ success probability is achieved already for $p=1$. If, instead, the initial state is an eigenstate of the mixer, $p=2$ is needed to reach $100\%$ success probability. A similar behavior was observed for the MaxCut problem in Ref.~\cite{govia_freedom_2021}, where it was shown that designing the mixer Hamiltonian to allow for rotations around the $XY$-axis leads to a performance increase. In the absence of single-photon losses, these success probabilities are well reproduced when simulating QAOA on cat qubits. Deviations from the ideal case still arise, due to the imperfect average gate fidelities of the gates used to implement the sequence, as per \secref{sec:gates-cat-qubit}. In the presence of single-photon losses, the performances of the $R_X(\theta)$ and $R_Y(\theta)$ mixers are almost the same.

\begin{table}[!b]
    \caption{\label{tab:results}Performance of QAOA for solving a toy two-qubit instance of Exact Cover on cat qubits for different mixers and initial states. The percentages correspond to the success probability given by Eq.~\eqref{eq:fidelity}, using the optimal angles found numerically. The noisy case corresponds to a single-photon loss rate of $K/1500$. The simulation results were obtained by solving the Schrödinger equation for the ideal case with no losses, and the Lindblad master equation for the noisy case.}
    \begin{ruledtabular}
    \begin{tabular}{c c c c c c}
        $p$ & Input & Mixer & \makecell[c]{Ideal \\ QAOA (\%)} & \makecell[c]{Cat qb. with \\ no losses (\%)} & \makecell[c]{Cat qb. with \\ losses (\%)} \\ \midrule
         1 & $\ket{+}$ & $X$ & 50 & 50.0 & 49.0 \\
         2 & $\ket{+}$ & $X$ & 100 & 99.9 & 90.6 \\
         1 & $\ket{+i}$ & $X$ & 100 & 99.9 & 96.4 \\
         1 & $\ket{+i}$ & $Y$ & 50 & 49.9 & 48.4 \\
         2 & $\ket{+i}$ & $Y$ & 100 & 99.9 & 91.3 \\
         1 & $\ket{+}$ & $Y$ & 100 & 99.9 & 95.8 \\
    \end{tabular}
    \end{ruledtabular}
\end{table}

\subsection{\label{sec:Pauli-transfer}Numerical results for larger systems: Pauli transfer matrix formalism}
We now move forward to more complex simulations. In this section, we numerically simulate solving 6, 8, and 10-qubit MaxCut problems using QAOA with cat qubits and compare it to the case of standard qubits, encoded in two-level systems, in the presence of single-photon loss for both systems. The MaxCut problem is an NP-complete problem that has been extensively studied in the context of QAOA~\cite{Farhi2020,Wurtz2020,arute2020quantum}. The objective of MaxCut is to partition the set of vertices of a graph into two subsets, such that the sum of the edge weights going from one partition to the other is maximum. MaxCut can be formulated as follows: Given a graph $G=(V,E)$, where $V$ is the set of vertices, and $E$ is the set of edges, the MaxCut Hamiltonian is
\begin{equation}
    \hat H_C=\frac{1}{2}\sum_{i,j\in E}(1-\hat Z_i\hat Z_j),
\end{equation}
where the sum is over all edges.

Since the total Hilbert space dimension increases exponentially with the number of KNRs --- The Hilbert space for each KNR is infinite but we truncate it in our numerical simulations to 20 photon levels for each resonator --- simulating more than 2 to 3 KNRs quickly becomes computationally difficult. A different strategy is to perform quantum gate set tomography by using the Pauli transfer matrix (PTM) formalism. This technique allows us to map the quantum process of each individual gate to effective two-level systems and, hence, simulate it using the Kraus-operator formalism instead of the Lindblad master equation, which is a lot more computationally efficient.
\begin{table*}[tbp]
    \caption{\label{tab:errorchannel}Error channel of the cat qubit for the $R_Z(\pi/2)$, $R_X(\pi/2)$, and $R_{ZZ}(\pi/2)$-gate. The error channel is shown for coefficients greater than $\geq 10^{-3}$.}
    \begin{ruledtabular}
    \begin{tabular}{l c c}
        Gate & Error Channel & Coefficients \\
        \midrule
        $R_Z(\pi/2)$ & $(1-p)\rho + p Z\rho Z$  & $p = 0.006$ \\[.5em]
        $R_X(\pi/2)$ & $(1-p_Y-p_Z)\rho + p_Z Z\rho Z + p_Y Y\rho Y + p_{YZ} Y\rho Z + p_{ZY} Z\rho Y$ & $p_Y = 0.01$, $p_Z=0.01$, $p_{YZ}=p_{ZY}=-0.002$  \\[.5em]
        $R_{ZZ}(\pi/2)$ & $(1-p_{1}-p_{2})\rho + p_{1} Z_1\rho Z_1+ p_{2} Z_2\rho Z_2$ & $p_{1} = 0.005$, $p_{2}=0.005$ \\
    \end{tabular}
    \end{ruledtabular}
\end{table*}

For a quantum channel $\mathcal{E}(\rho)$ the PTM is formally defined as~\cite{greenbaum_tomography_2015}
\begin{equation}
    (R_\mathcal{E})_{ij}\equiv\frac{1}{d}\Tr[\hat P_i\mathcal{E}(\hat P_j)],
\end{equation}
where $\hat P_j\in\{\hat I,\hat X,\hat Y,\hat Z\}^{\otimes n}$ is the Pauli group in the computational basis for $n$-qubits, and $d=2^n$ is the Hilbert space dimension. Furthermore, the PTM formalism allows for composite maps to be written as a matrix product of the individual PTMs, i.e. $\mathcal{E}_2\circ\mathcal{E}_1 = R_{\mathcal{E}_2}R_{\mathcal{E}_1}$. Using this fact, we can deconstruct the PTM as a product of two parts: an ideal part $R_\mathrm{ideal}$, corresponding to the noiseless ideal gate, and a noise part $R_\mathrm{noise}$, corresponding to both coherent errors as a result of imprecise unitary operation, and incoherent errors stemming from single-photon losses. Since the ideal gate operation is known, it is possible to extract the erroneous part from the full quantum process as follows:
\begin{equation}
    R_{\mathcal{E}} = R_\mathrm{noise} R_\mathrm{ideal} \Rightarrow R_\mathrm{noise} = R_{\mathcal{E}} R_\mathrm{ideal}^{-1}.
\end{equation}
We now use the aforementioned procedure in order to transform the continuous time evolution of the KNR gates to PTMs. Since the QAOA implementation of MaxCut only requires $R_X(\theta)$ and $R_{ZZ}(\Theta)$-gates, we will only focus on these two gates, starting with the former. Because the $R_X(\theta)$-gate is not noise bias preserving, meaning that single-photon losses do not commute through the gate, the noise part $R_\mathrm{noise}$ will ultimately depend on the angle $\theta$. We therefore compute $R_\mathrm{noise}$ for 180 evenly spaced points between $0$ and $\pi$ for the $R_X(\theta)$-gate, and use the closest $R_\mathrm{noise}$ for a given $\theta$ in upcoming simulations. Hence, we do not need to compute $R_{\mathcal{E}}$ for every possible angle. For the $R_{ZZ}(\Theta)$-gate, however, we only compute the PTM for $\Theta=0$, since this gate is noise bias preserving. That is, a single-photon loss corresponds to a $\hat Z$ error in the computational subspace, and $R_\mathrm{noise}$ is thus independent on the angle $\Theta$. For the MaxCut problem, the $R_Z(\phi)$-gate is not needed for the circuit compilation, and we therefore exclude it.

Once the PTMs have been obtained, we transform them to Kraus operators in order to easily simulate the circuit using Cirq~\cite{cirq_developers_2021} as
\begin{equation}
    \hat \rho \rightarrow \sum_{k=1}^m \hat A_k (\hat U\hat \rho \hat U^\dagger)\hat A_k^\dagger,
\end{equation}
where $\hat U$ corresponds to the ideal gate and $\hat A_k$ is the set of Kraus operators that describes the noise. Transforming the PTM to Kraus operators can be done by first transforming the PTM to the Choi representation and then transform the Choi-representation to the Kraus representation. To begin, the PTM for a $n$-qubit channel can be transformed to a Choi-matrix according to~\cite{greenbaum_tomography_2015}
\begin{equation}
    \hat \rho_\mathcal{E} = \frac{1}{d^2}\sum_{i,j=1}^{d^2}(R_\mathcal{E})_{ij}\hat P^T_j \otimes \hat P_i.
\end{equation}
Given the Choi-matrix, the Kraus-representation is obtained by first diagonalizing the Choi-matrix, from which its eigenvalues $\{\lambda_i\}$ and eigenvectors $\{|\hat A_i\rangle\rangle\}$ are extracted, where $|\cdot \rangle\rangle$ is a superoperator. The eigenvalues and eigenvectors are then used to construct the Kraus operators as follows~\cite{wood_tensor_2015}:
\begin{equation}
    \hat A_i=\sqrt{\lambda_i}\mathrm{unvec}(|\hat A_i\rangle\rangle),
\end{equation}
where $\mathrm{unvec}$ is the unvectorization operation. In \tabref{tab:errorchannel}, we present the error channels for the cat qubit for the $R_Z(\pi/2)$, $R_X(\pi/2)$, and $R_{ZZ}(\pi/2)$ gates, each expanded in the Pauli basis. The results clearly indicate that the $R_Z(\phi)$ and $R_{ZZ}(\Theta)$ gates preserve noise bias, while the $R_X(\theta)$ gate does not.
\begin{table}[!b]
    \centering
    \caption{Average gate fidelities for the $R_{ZZ}(\Theta)$ and $R_X(\theta)$-gate for cat qubits with $\alpha=1.36$ and standard qubits obtained using the Kraus operator formalism. The results are averaged over 20 points evenly spaced between 0 and $\pi$.}
    \begin{ruledtabular}
    \begin{tabular}{lcc}
    Avg. gate fid. (\%) & $R_{ZZ}(\Theta)$ & $R_X(\theta)$ \\ \midrule
    Cat qubits & 99.16     & 98.60 \\
    Standard qubits & 99.16 & 98.62 \\
    \end{tabular}
    \end{ruledtabular}
    \label{tab:fidelity}
\end{table}
\begin{figure*}[!t]
    \includegraphics[width=\linewidth]{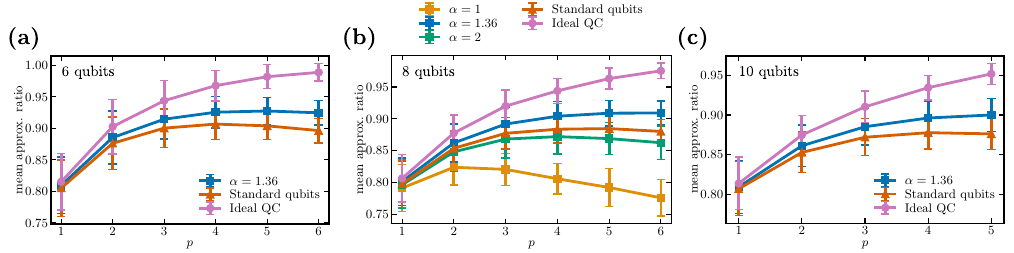}
    \caption{\label{fig:max-cut-results}Mean approximation ratio averaged over 30 instances for MaxCut graphs with 6, 8, and 10 qubits. For the 8-qubit case, we perform additional simulations for three coherent state amplitudes. The circle corresponds to the approximation ratio of an ideal (noise-free) quantum computer. The squares represent the approximation ratio obtained using cat qubits, while the triangles denote the ratio with standard qubits encoded into discrete two-level systems. The average gate fidelity for both cat qubits with amplitude $\alpha=1.36$ and standard qubits was nearly identical, with values reported in \mbox{\tabref{tab:fidelity}}.}
\end{figure*}

In order to make a fair comparison between the performance of the cat qubit and the one of the standard qubit, we chose the relevant parameter such that the average gate fidelities are the same between the two systems. By doing so, we can compare which encoding, continuous versus discrete, is the best for QAOA. For the standard qubit device, we implement the $R_X(\theta)$-gate by evolving under the Pauli $\hat X$, and the $R_{ZZ}(\Theta)$-gate by evolving under $\hat Z_i\hat Z_j$. The gate time $T_g$ is chosen to be the same as was used for the cat qubit device, i.e. $T_g=10/K$ where $K$ is the Kerr non-linearity for the $R_X(\theta)$-gate and $T_g=2/K$ for the $R_{ZZ}(\Theta)$-gate. We specifically pick the relaxation rates $T_1$ with the pure dephasing rate $T_\phi$ set to zero, such that the average gate fidelity corresponds to that of the KNR-gates for $\alpha=1.36$, which maximizes the average gate fidelity in Fig.\ref{fig:rz-alpha}. To this aim, we use an expression for the first-order reduction in the average gate fidelity due to relaxation rate~\cite{abad2021universal}
\begin{equation}
    \label{eq:gate-fidelity}
    \bar F = 1 - \frac{d}{2(d+1)} T_g n \Gamma_1,
\end{equation}
where $\bar F$ is the average gate-fidelity, $d=2^n$, and $\Gamma_1=1/T_1$ is the relaxation rate where $T_1$ is the relaxation time which we assume to be the same for all $n$ qubits. The expression can be re-written to give the relaxation rate in terms of the average gate-fidelity
\begin{equation}
    \label{eq:relaxation}
    \Gamma_1 = 2\frac{(d+1)(1-\bar F)}{dT_g n}.
\end{equation}
Using the average gate-fidelities $\bar F$ that were numerically calculated for the cat qubits in \tabref{tab:avg-fidelity}, the corresponding relaxation rates for the standard qubits that results in the same average gate fidelity as for the cat qubits can be obtained.

Likewise, we do quantum gate set tomography using the PTM formalism for the standard qubit device, revealing a less structured error channel as compared as the cat qubit case. Since neither the $R_X(\theta)$ nor the $R_{ZZ}(\Theta)$-gate are noise bias preserving in this case, we compute $R_\mathrm{noise}$ for 180 evenly spaced points of $\theta$ and $\Theta$ between $0$ and $\pi$ for each of the two gates respectively. In \tabref{tab:fidelity} we report the average gate fidelity for the $R_{ZZ}(\Theta)$ and $R_X(\theta)$-gate using the Kraus operator formalism for both cat qubits and the standard qubits after setting the relaxation time $T_1$ found for the standard qubits. From \tabref{tab:fidelity} the average gate fidelities match very well between the cat and standard qubits, with the $R_X(\theta)$-gate being $0.02\%$ higher for the standard qubits, which we attribute to the fact that Eq.~\eqref{eq:gate-fidelity} is only a first-order approximation of the average gate fidelity.

Using the Kraus-operator formalism, we are able to simulate QAOA with cat qubits and standard qubits for solving $30$ randomly generated 6, 8 and 10-qubit instances of MaxCut on Erdős--Rényi graphs with edge probability $p=0.5$. As a metric for comparison between the performance of cat qubits and standard qubits, we look at the approximation ratio, defined as
\begin{equation}
    r\equiv\frac{\Tr(\hat \rho \hat H_C)}{C_\text{max}},
\end{equation}
where the numerator is the expected cut value with $\hat \rho$ the density matrix output from QAOA, and $C_\text{max}$ is the value of the maximum cut. The simulation results are presented in \figref{fig:max-cut-results}. For both standard and cat-qubits, the approximation ratio first increases at increasing $p$, and then starts decreasing when $p$ is sufficiently high so that the noise in the gates for implementing the QAOA sequence makes it less advantageous to use large depth circuits. The results show that given the same average gate fidelities, the approximation ratio obtained for the KNR device is higher than for the standard qubit device for all iteration levels $p$, thereby indicating an advantage in the use of the former qubit implementation over the latter. The numerical method used to achieve the classical optimization of the various QAOA instances is described in \appref{app:approxr}, where we also report on the simulation results for the approximation ratios corresponding to the ideal case, the use of cat qubits, and of standard qubits respectively, without averaging over the random instances.

We provide an explanation for the better performance of cat qubits, by looking at how errors are transformed during the $R_X(\theta)$ and $R_{ZZ}(\Theta)$ gates. As discussed, the dominant error mechanism, i.e. single-photon losses, translates into $\hat{Z}$-errors in the cat qubit. Consider a $\hat Z$ error occurring before the $R_X(\theta)$ gate. By commuting this error through the gate, we derive the following:
\begin{align}
    R_X(\theta)\hat Z &= e^{-i \theta \hat X/2}\hat Z \notag\\
    &= \cos(\frac{\theta}{2})\hat Z-i\sin(\frac{\theta}{2})\hat X\hat Z
    \notag\\
    &= \hat Z R_X(-\theta),
\end{align}
where in the final equality we have used the anti-commutation relation $\hat X\hat Z=-\hat Z\hat X$. This shows that a $\hat Z$ error effectively introduces an extra $\pi$ rotation in the $R_X(\theta)$ gate. Conversely, for the $R_{ZZ}(\Theta)$ gate, a $\hat Z$ error on either qubit commutes through the gate. 

\mbox{\tabref{tab:errorchannel}} highlights the error channel affecting the cat qubit for these two gates, showing that while the $R_{ZZ}(\Theta)$ gate retains its noise channel structure post-gate (with $\hat Z$ errors), the $R_X(\theta)$ gate alters the channel, introducing $\hat Y$ components. In other words, the $R_{ZZ}(\Theta)$ gate is noise-bias preserving while $R_X(\theta)$ is not.

Conversely, one can commute the error through the gate, while preserving the structure of the gate. For the example above, this results in:
\begin{equation}
    R_X(\theta)\hat Z = \qty(\cos(\theta)\hat Z -\sin(\theta) \hat Y)R_X(\theta).
\end{equation}
This shows that the $R_X(\theta)$-gate transforms the $\hat Z$ error into a combination of $\hat Z$ and $\hat Y$ errors, and, in other words, it makes the noise channel unbiased, as can be seen from \mbox{\tabref{tab:errorchannel}}.

Moreover, for a typical MaxCut graph of level-$p$, the number of $R_{ZZ}(\Theta)$ gates corresponds to the graph's edge count, which averaged 14.1 for the 8-qubit instances, while the count of $R_X(\theta)$ gates is equal to the number of vertices or qubits. There are, in other words more, $R_{ZZ}(\Theta)$ gates than $R_{X}(\Theta)$ gates in a typical QAOA circuit for the Max-cut problems considered. In contrast to cat qubits, standard qubits are susceptible to $\hat Z$, $\hat X$, and $\hat Y$ errors, affecting the fidelity of both $R_X(\theta)$ and $R_{ZZ}(\Theta)$ gates.

We note that the approximation ratio for both standard and cat qubits could be further improved by resorting to error mitigation techniques for estimating expectation values. In particular, in the specific case of virtual distillation~\mbox{\cite{cotler_quantum_2019,koczor_exponential_2021,huggins_virtual_2021}}, when noise is present during the error mitigation procedure itself, the advantage of using cat qubits is further enhanced over the use of standard, non noise-biased qubits~\mbox{\cite{vikstal_study_2022}}. Finally, one might wonder whether a better performance in terms of the approximation ratio could be obtained by defining a genuinely bosonic variant of the QAOA algorithm, i.e. by Trotterizing an appropriate bosonic quantum annealing Hamiltonian \cite{puri_annealing_2017}, which initial optimal eigenstate is the vacuum, and which final optimal eigenstate encodes the solution. We explore this question in Appendix \ref{app:bosonic QAOA}, where we compare the needed mixing and cost Hamiltonian to the case of standard QAOA implemented with cat qubits.  Our numerical simulations suggest that such a bosonic QAOA algorithm does not yield an improvement over QAOA on cat qubits. We hypothesize that this missed efficiency stems from the need for bosonic QAOA to bring the state of the system of resonators into the qubit computational basis.

\section{\label{sec:Conclusion}Conclusions}
In conclusion, we have studied implementations of QAOA with a noise-biased qubit, namely the cat qubit, and we have performed numerical simulations in the case that such a cat qubit is implemented by means of a  Kerr nonlinear resonator. Despite the algorithmic sequence requiring non-bias preserving $X$-rotations, running QAOA on such cat qubits yields a performance advantage with respect to the use of standard qubits in the presence of noise caused by single-photon losses for the studied problem, MaxCut. We expect these results to not be dependent on the chosen problem, and that other problems than MaxCut would benefit from the same performance separation. 

Our results indicate that noise-biased qubits that favor dephasing errors, such as cat qubits, are preferable over standard qubits for the implementation of QAOA on near-term intermediate-scale quantum processors, and provide a concrete estimate of the obtainable approximation ratio for MaxCut, for an implementation based on Kerr resonators with realistic noise parameters.

One challenge that pertains to variational algorithms is the so-called barren plateau problem, which refers to the lack of trainability of the parameters caused by an exponentially flat cost function landscape, see e.g. \mbox{\cite{bittel2021training}}. However, works that consider suboptimal choices of parameters \cite{sreedhar2022quantumapproximateoptimizationalgorithm, montanezbarrera2024universalqaoaprotocolevidence} yield viable solutions to the optimization for sizes of up to about hundred qubits. In particular, Ref.~\cite{montanezbarrera2024universalqaoaprotocolevidence} points to the possibility of using linear ramps in combination with QAOA (LR-QAOA), where the parameters are not globally optimised, but instead, linear interpolation is considered. An interesting direction would be to tackle larger system sizes, which would require deeper circuits, where making use of these strategies might be helpful.

A further interesting question that stems from our work is if the results here presented, and in particular the performance of QAOA with cat qubits, would further improve in the case where one would adopt a similar encoding of cat qubits in Kerr resonators, but with a more sophisticated use of the detuning as was introduced in Ref.~\cite{Ruiz_twophoton_2022}, or with the dissipative cat qubits and corresponding gates considered in Ref.~\cite{Gautier2023}. We leave this analysis for future work.

After the online submission of our work as an arXiv preprint we came across an independent work on QAOA with cat qubits~\cite{bornens2023variational}.

%
\acknowledgments
We acknowledge useful discussions with Simone Gasparinetti and Timo Hillmann, as well as Laurent Prost for having suggested us to optimise  the amplitude of the coherent states. G. F. acknowledges support from the Vetenskapsrådet (Swedish Research Council) Grant QuACVA. G. F., L. G.-\'{A}., and P. V. acknowledge support from the Knut and Alice Wallenberg Foundation through the Wallenberg Center for Quantum Technology (WACQT). S. P. was supported by the Air Force Office of Scientific Research under award number FA9550-21-1-0209.

\codeavailability
\indent The code used for producing the results is made available in Ref.~\cite{pontus_2023}. All master equation simulations are performed using QuTip~\cite{johansson_qutip_2012,johansson_qutip_2013,li_qip_2022}.

\appendix
\section{\label{app:avg-gate-fid}Quantum gates on cat qubits}
In this Appendix, we will go through the implementation of a universal gate set for the cat qubits implemented in a Kerr nonlinear resonator. All gates will be evaluated in terms of their average gate fidelity. The average gate fidelity of a quantum channel $\mathcal{E}$ for a qudit of dimension $d$ is defined as~\cite{Nielsen2002}
\begin{equation}
    \label{eq:avg-gate-fid-qudit}
    \bar F(\mathcal{E},\hat U)=\frac{\sum_j\Tr(\hat U\hat P_j^\dagger \hat U^\dagger \mathcal{E}(\hat P_j))+d^2}{d^2(d+1)},
\end{equation}
where $\hat U$ is the target gate and the sum is over the basis of unitary operators $\hat P_j$ for the qudit, with $\hat P_j$ satisfying $\Tr(\hat P_j^\dagger \hat P_k)=\delta_{jk}d$. In the simulations we set $d=2$ for single-qubit gates and $d=4$ for two-qubit gates, and $\hat P_j$ is chosen to be one of the Pauli matrices in the computational basis, e.g $\hat P_j\in\{\hat I,\hat X,\hat Y,\hat Z\}^{\otimes n}$, where $n$ is the number of cat qubits. Moreover, in all subsequent simulations we set $\alpha=1.36$, which corresponds the coherent state amplitude with the highest gate fidelity.
%
\subsection{\label{sec:rz-gate}\texorpdfstring{$R_Z(\phi)$}{RZ}-gate}
\begin{figure}[htbp]
    \centering
    \includegraphics[width=\linewidth]{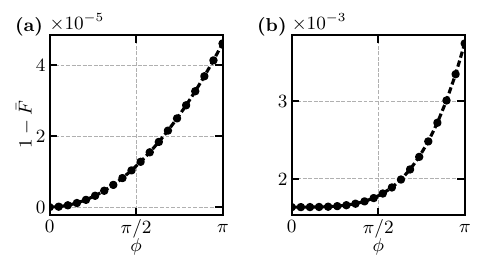}
    \caption{The average gate infidelity $(1-\bar F)$ of the $R_Z(\phi)$-gate\textbf{(a)} without noise and \textbf{(b)} with a single-photon loss rate of $K/1500$.}
    \label{fig:rz}
\end{figure}

The $R_Z(\phi)$ gate can be performed by applying a single-photon drive with an amplitude of $E(t)$ to the KNR. The Hamiltonian for this drive is given by:
\begin{equation}
\label{eq:single-photon-drive}
    \hat H_Z(t) = E(t)(\hat ae^{-i\theta} + \hat a^\dagger e^{i\theta}),
\end{equation}
where $\theta$ is the phase of the drive. When $\theta=0$, and $\abs{E(t)}\ll 4G$ and the variation of $E(t)$ is sufficiently slow, the cat qubit is approximately kept in the computational basis~\cite{goto_universal_2016,Vikstal2018}. Applying the projector onto the computational subspace $\hat I = \dyad{\bar 0} + \dyad{\bar 1}$ to the single-photon drive Hamiltonian Eq.~\eqref{eq:single-photon-drive} gives for large $\alpha$
\begin{equation}
    \hat I E(t)(\hat a^\dagger + \hat a)\hat I = 2E(t)\alpha\hat Z.
 \end{equation}
We perform numerical simulations where we set $\Delta = 0$ and define $E(t)$ as
\begin{equation}
    \label{eq:SPD-Amplitude}
    E(t) = \frac{\pi \phi}{8T_g\alpha}\sin\frac{\pi t}{T_g},
\end{equation}
with $T_g = 2/K$, and $\phi$ is the angle for the gate. In \figref{fig:rz} the average gate infidelity $(1-\bar F)$ as a function of $\phi$ for the $R_Z(\phi)$-gate is shown: in \textbf{(a)} without losses, and in \textbf{(b)} with a single-photon loss rate of $K/1500$.
\subsection{\label{sec:rx-gate}\texorpdfstring{$R_X(\theta)$}{RX}-gate}
\begin{figure}[tbp]
    \includegraphics[width=\linewidth]{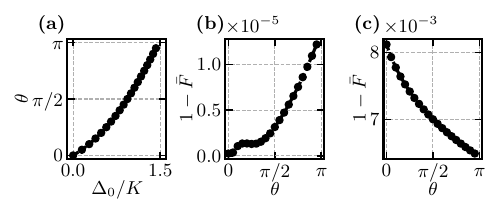}
    \caption{\textbf{(a)} $\theta$ maximizing the average gate fidelity $1-\bar F$ as a function of $\Delta_0$. \textbf{(b)} Average gate infidelity without noise and \textbf{(c)} with single-photon loss rate of $K/1500$.}
    \label{fig:rx}
\end{figure}

An $R_X(\theta)$-gate can be realized by means of a small non-zero detuning $\Delta$ between the two-photon drive and the resonator. This can be understood by projecting the number operator in the computational basis:
\begin{align*}
    \hat I \hat a^\dagger \hat a \hat I = |\alpha|^2 \hat I - |\alpha|^2e^{-2|\alpha|^2}\hat X.
\end{align*}
If $\Delta(t)\ll 2G$ the computational states $\ket{\bar 0}$ and $\ket{\bar 1}$ are approximately kept in the computational subspace. Thus choosing $\Delta(t)$ as
\begin{equation}
    \Delta(t) =\frac{\theta\pi}{4T_g|\alpha|^2e^{-2|\alpha|^2}}\sin\frac{\pi t}{T_g}
\end{equation}
yields
\begin{equation}
    e^{-i[-\hat a^\dagger \hat a \int_0^{T_g}\Delta(t)dt]} = e^{-i\frac{\theta}{2}\hat X},
\end{equation}
corresponding to a $R_X(\theta)$-gate. The disadvantage of this approach, however, is that the gate time $T_g$ has to be exponentially large with respect to $\alpha$ in order to satisfy the condition $\Delta(t)\ll 2G$. For example, if $\alpha=2$, then a total gate time $T_g>1000/K$ is required. However, a second proposal was put forward by Goto~\cite{goto_universal_2016}, where the detuning is set to a fixed value $\Delta_0$, and the corresponding $\theta$ that maximizes the average gate fidelity is evaluated. Hence, to perform the $R_X(\theta)$-gate, $\Delta(t)$ is set to
\begin{equation}
    \label{eq:Delta}
    \Delta(t) = \Delta_0 \sin^2\frac{\pi t}{T_g},
\end{equation}
with $T_g=10/K$. Throughout this paper, we use this second method. We find the $\theta$ that maximizes the average gate fidelity for 20 values of $\Delta_0$ between $0$ and $3.95 K$, see \figref[a]{fig:rx}. It can be seen that while $\Delta_0$ changes from $0$ to $3.95 K$, the rotation angle $\theta$ changes from $0$ to $\pi$. In \figref{fig:rx} \textbf{(b)} and \textbf{(c)} the average gate infidelity ($1-\bar F$) as a function of $\theta$ for the $R_X(\theta)$-gate is shown: in \textbf{(b)} without losses, and in \textbf{(c)} with a single-photon loss rate of $K/1500$.
\subsection{\label{sec:ry-gate}\texorpdfstring{$R_Y(\varphi)$}{RY}-gate}
\begin{figure}
    \includegraphics{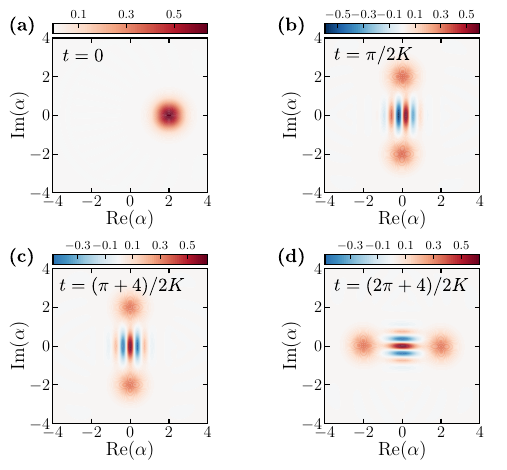}
    \caption{\textbf{(a)}-\textbf{(d)} Wigner function at four different stages of the $R_Y(\pi/2)$-gate starting from the $\ket{\bar 0}$ state. Between \textbf{(a)}-\textbf{(b)} the two-photon drive is turned off to let the state evolve freely under the Kerr Hamiltonian. When a time $t=\pi/2K$ has passed, the two-photon drive is turned on again but this time with a $\pi/2$ phase such that the state is stabilized along the imaginary axis in the phase space. Between \textbf{(b)}-\textbf{(c)}, a single-photon drive with a $\pi/2$ is applied to the cat-state for a time $t=2\pi/K$. This makes the superposition of the two coherent states acquire a phase difference depending on the angle $\varphi$. Finally, between \textbf{(c)}-\textbf{(d)}, the two-photon drive is turned off once more for a time $t=\pi/2K$ to let the state evolve back and be stabilized along the real axis.}
    \label{fig:ry-wigner}
\end{figure}

\begin{figure}[htbp]
    \includegraphics[width=\linewidth]{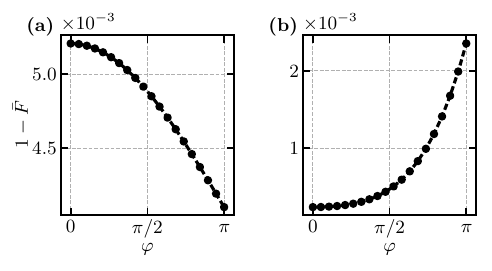}
    \caption{The average gate infidelity $1-\bar F$ of the $R_Y(\varphi)$-gate \textbf{(a)} without noise and \textbf{(b)} with single-photon loss rate of $K/1500$.}
    \label{fig:ry}
\end{figure}

To perform the $R_Y(\varphi)$-gate, the two-photon drive is turned off for a total time $t=\pi/2K$ to let the state evolve freely under the Kerr Hamiltonian. If the initial state is the vacuum state $\ket{0}_\mathrm{vac}$, it will evolve into $(\ket*{C^+_{i\alpha}}+i\ket*{C^-_{-i\alpha}})/\sqrt{2}$. Once the state is along the imaginary axis, the two-photon drive is turned on, with a $\pi/2$ phase, so that the state is stabilized along the imaginary axis. Applying the single-photon drive also with a $\pi/2$ phase, such that $\hat H_Z(t) = E(t)(\hat a^\dagger e^{i\pi/2} + \hat ae^{-i\pi/2})$, where $E(t)$ is given by Eq.~\eqref{eq:SPD-Amplitude}, the two cat states will acquire a phase difference. When the two-photon drive is turned off for a second time, that is $t=\pi/2K$, the resulting gate is $R_Y(\varphi)$, see \figref{fig:ry-wigner}. In \figref{fig:ry} the average gate infidelity $(1-\bar F)$ as a function of $\varphi$ for the $R_Y(\varphi)$-gate is shown in \textbf{(a)} without losses, and with a single-photon loss rate of $K/1500$ in \textbf{(b)}. 
\subsection{\label{sec:U-gate}\texorpdfstring{$R_{ZZ}(\Theta)$}{RZZ}-gate}
\begin{figure}
    \includegraphics[width=\linewidth]{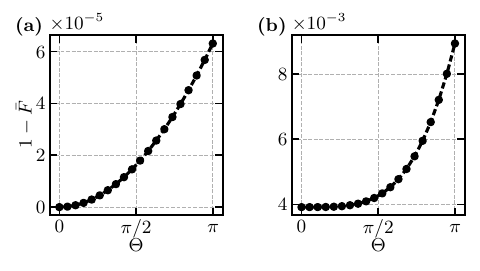}
    \caption{The average gate infidelity $1-\bar F$ of the $R_{ZZ}(\Theta)$-gate \textbf{(a)} without noise and \textbf{(b)} with single-photon loss rate of $K/1500$.}
    \label{fig:rzz}
\end{figure}

The two-qubit Ising-zz gate $R_{ZZ}(\Theta)$ is achieved by means of two-photon exchange between two KNRs, yielding the coupling Hamiltonian 
\begin{equation}
\label{eq:coupling-Hamiltonian}
    \hat H_{ZZ} = g(t)(\hat a_1 \hat a_2^\dagger + \hat a_1^\dagger \hat a_2).
\end{equation}
When $|g(t)|\ll 2G$, the KNRs are approximately kept in the subspace spanned by $\ket{\bar 0 \bar 0}$, $\ket{\bar 0\bar 1}$, $\ket{\bar 1\bar 0}$ and $\ket{\bar 1\bar 1}$. Projection of Eq.~\eqref{eq:coupling-Hamiltonian} onto the computational basis yields for large $\alpha$
\begin{equation}
	\hat H_{ZZ}
    = 2\alpha^2g(t)\hat{Z}_1\hat{Z}_2 +\text{const}.
\end{equation}
In our numerical simulation we set $T_g=2/K$, and to perform $R_{ZZ}(\Theta)$, we set $g(t)$ as
\begin{equation}
    g(t) = \frac{\pi\Theta}{8T_g\alpha^2}\sin\frac{\pi t}{T_g}.
\end{equation}
In \figref{fig:rzz} the average gate infidelity $(1-\bar F)$ as a function of $\Theta$ for the $R_{ZZ}(\Theta)$-gate is shown: in \textbf{(a)} without losses, and  in \textbf{(b)} with a single-photon loss rate of $K/1500$.
\section{\label{app:approxr}Numerical optimization and approximation ratios}
\begin{figure}[htp]
    \centering
    \includegraphics{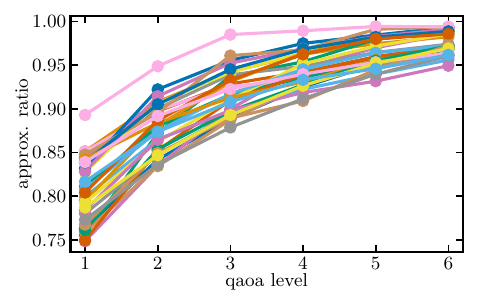}
    \caption{The approximation ratio as a function of the QAOA level $p$ plotted for each individual instance in the ideal case, meaning no noise. There are 30 instances in total.}
    \label{fig:approxr_ideal}
\end{figure}

In this section we elaborate on the classical optimization part of QAOA that was used in \secref{sec:Pauli-transfer}. For $p=1$, brute force optimization is used where the cost function $\matrixelement{\psi_1(\gamma,\beta)}{\hat H_C}{\psi_1(\gamma,\beta)}$ is evaluated on a $100\times 100$ grid. For $p>1$, we use the interpolation method, described in Ref.~\cite{zhou_quantum_2020}, together with a local optimizer. This strategy consists in predicting a good starting point for the variational parameters search at level $p+1$ for each individual instance based on the best variational parameters found at level $p$ for the same instance. From the produced starting point, we run a \textsc{L-BFGS} optimizer. \figref{fig:approxr_ideal} shows the approximation ratio for each instance for noiseless, ideal QAOA as a function of the level $p$. As can be seen from the figure, the approximation ratio increases at increasing QAOA level for each individual instance, indicating the success of the classical optimizer at finding good variational parameters.
\section{\label{app:bosonic QAOA}Bosonic QAOA}
In this Appendix we explore the possibility of deriving a genuinely bosonic version of the QAOA algorithm from Trotterizing a bosonic quantum annealing Hamiltonian, in analogy to what was initially done for qubit QAOA in Ref.~\cite{farhi2014quantum}. We refer to this new algorithm as Bosonic QAOA. We compare numerically its performance to QAOA on cat qubits, for the simple case of finding the ground state of a single Ising spin. In this Appendix we chose a coherent state amplitude of $\alpha = 2$~\mbox{\cite{puri_engineering_2017}}.

\subsection{\label{app:bosonic QAOA-Trotter}Trotterization of the CV Quantum Annealing Hamiltonian}
\begin{table*}[tbp]
    \caption{\label{tab:mixing-hamiltonian}Comparison between bosonic QAOA and QAOA in terms of mixer Hamiltonian, cost Hamiltonian and input state.}
    \begin{ruledtabular}
    \begin{tabular}{l l l}
        & bosonic QAOA & Standard QAOA
        \\
        \midrule
        $\hat H_M$
        & $\sum\limits_{i=1}^n[-\Delta \hat a_i^\dagger \hat a_i - K\hat a_i^{\dagger 2}\hat a_i^2]$ 
        & $\sum\limits_{i=1}^n[-\Delta(t) \hat a_i^\dagger \hat a_i - K\hat a_i^{\dagger 2}\hat a_i^2 + G(\hat a_i^{\dagger 2} + \hat a_i^2)]$
        \\[1.5em]
        $\hat H_C$
        & $\begin{array}{l}
             \sum\limits_{i=1}^n [- K\hat a_i^{\dagger 2}\hat a_i^2 +G(\hat a_i^{\dagger 2}+\hat a_i^2) + E_i(\hat a_i+\hat a_i)] \\
             + \sum\limits_{1\leq i<j\leq n} g_{ij}(\hat a_i^\dagger \hat a_j + \hat a_j^\dagger \hat a_i)
        \end{array}$
        & $\begin{array}{l}
             \sum\limits_{i=1}^n[-K\hat a_i^{\dagger 2}\hat a_i^2 + G(\hat a_i^{\dagger 2} + \hat a_i^2) + E_i(t)(\hat a_i+\hat a_i)]] \\ 
             + \sum\limits_{1\leq i<j\leq n} g_{ij}(t)(\hat a_j^\dagger\hat a_i + \hat a_i^\dagger \hat a_j)
        \end{array}$
        \\[2.5em]
        Input & $\ket{0}_\mathrm{vac}$ & $\ket{C^+_\alpha}$
        \\
    \end{tabular}
    \end{ruledtabular}
\end{table*}
The time evolution of the quantum annealing algorithm starts from the ground state of a Hamiltonian that is easy to prepare and slowly evolves the system into the ground state of a Hamiltonian encoding the solution to a combinatorial optimization problem. If the evolution is slow enough, as set by the quantum adiabatic theorem, the initial state will follow the instantaneous ground state throughout the evolution and end up in the solution state. The algorithm also works if the initial state is the highest energy eigenstate, or ``roof'' state, of the initial Hamiltonian, provided the final Hamiltonian encodes the solution in its highest energy eigenstate.

The starting point for deriving a bosonic QAOA algorithm is the quantum annealing Hamiltonian
\begin{align}
    \label{eq:RotatingFrame}
	\hat{H}(t)
    = \left(1-\frac{t}{\tau}\right) \hat{H}_M
    +\frac{t}{\tau}\hat{H}_C,
\end{align}
where $\hat H_M$ is the initial Hamiltonian, whose ground state is easy to prepare, and $\hat H_C$ is the final Hamiltonian, whose ground state encodes the solution to an optimization problem. We take inspiration from the annealing protocol using Kerr resonators of Ref.~\cite{puri_annealing_2017}. For $n$ resonators, we can choose
\begin{equation}
    \label{eq:trotter-mixing-hamiltonian}
    \hat H_M = 
    \sum_{i=1}^n(-\Delta \hat a_i^\dagger \hat a_i - K\hat a_i^{\dagger 2}\hat a_i^2),
\end{equation}
which has the vacuum state $\ket{0}_\mathrm{vac}$ as its ``roof state'', and
\begin{multline}
    \label{eq:trotter-cost-hamiltonian}
    \hat H_C = \sum_{i=1}^n \Big[ - K\hat a_i^{\dagger 2}\hat a_i^2 +G\qty(\hat a_i^{\dagger 2}+\hat a_i^2) \\
    + E_i\qty(\hat a_i^\dagger + \hat a_i)\Big] 
     + \sum_{1\leq i<j\leq n} g_{ij}\qty(\hat a_i^\dagger \hat a_j + \hat a_j^\dagger \hat a_i).
\end{multline}
By starting from the vacuum state and slowly increasing $t$, the instantaneous eigenstate of the Hamiltonian Eq.~\eqref{eq:RotatingFrame} evolves into the highest energy eigenstate of $\hat H_C$ which encodes the solution to an optimization problem upon cat qubit encoding~\cite{puri_annealing_2017}.

We will now, in the spirit of Farhi et al.~\cite{farhi2014quantum}, Trotterize the bosonic quantum annealing Hamiltonian Eq.~\eqref{eq:RotatingFrame} to obtain a genuinely bosonic version of QAOA. The continuous-time evolution governed by the time-dependent Hamiltonian of Eq.~\eqref{eq:RotatingFrame} is given by
\begin{align}
    \label{eq:time-evolution-operator}
    \hat U(T) &\equiv\mathcal{T}\exp[-i\int_0^T \hat H(t) dt] \notag \\
    &\approx\prod_{k=1}^{p}\exp[-i \hat H(k\delta t) \delta t],
\end{align}
where $\hat U(T)$ is the evolution operator from $0$ to $T$, $\mathcal{T}$ is the time-ordering operator, and $p$ is a large integer so that $\delta t=T/p$ is a small time interval. Since $\hat H_M$ and $\hat H_C$ are two non-commuting Hamiltonians, one can use the Trotter formula:
\begin{equation}
    e^{i (A+B)\delta t} = e^{i A\delta t} e^{i B\delta t} + \mathcal{O}(\delta t^2),
\end{equation}
for two non-commuting operators $A$ and $B$ given sufficiently small $\delta t$, and apply it to the discretized time evolution operator Eq.~\eqref{eq:time-evolution-operator}, yielding
\begin{multline}
    \hat U(T) \approx
    \prod_{k=1}^{p}\exp[-i \qty(1-\frac{k\delta t}{\tau})\hat H_M \delta t]\\
    \times \exp[-i \frac{k\delta t}{\tau}\hat H_C \delta t].
\end{multline}
We have so far approximated the continuous-time evolution by a sequential product of discrete time steps. We can now apply the same idea underlying the QAOA algorithm in Ref.~\cite{farhi2014quantum}, which consists in truncating this product to an arbitrary positive integer $p$ and redefining the time dependence in each exponent in terms of variational parameters $\qty(1-k\delta t/\tau)\delta t\rightarrow\beta_k$ and $(k\delta t/\tau)\delta t\rightarrow\gamma_k$, leading to
\begin{equation}
    \label{eq:cvQAOA}
    \hat U_p =
    \prod_{k=1}^{p}\exp[-i \beta_k\hat H_M] \exp[-i \gamma_k\hat H_C].
\end{equation}
We then define our bosonic QAOA algorithm as the sequence in Eq.~\eqref{eq:cvQAOA}, with $\hat H_M$ and $\hat H_C$ given by Eq.~\eqref{eq:trotter-mixing-hamiltonian} and \eqref{eq:trotter-cost-hamiltonian} respectively, applied to vacuum state chosen as the initial state.

It is interesting to compare the bosonic QAOA algorithm that we derived to the standard QAOA from \secref{sec:QAOA-Kerr}, when the latter is implemented on cat qubits.

In \tabref{tab:mixing-hamiltonian} we compare the mixing Hamiltonian, cost Hamiltonian and initial states of bosonic QAOA and QAOA. Clearly, the cost Hamiltonian encoding the problem solution is the same for the two algorithms. Instead, the two-photon drive is not present in the mixer Hamiltonian for bosonic QAOA. The most notable difference is that while the input state for QAOA on cat qubits is the state $\ket{+}$ in all qubits, corresponding to initializing all qubits into a cat state $\ket{C^+_\alpha}$, the input state for bosonic QAOA is the vacuum state.
\subsection{\label{app:bosonic QAOA-numerics}Finding the ground state of a single Ising spin}
To test the performance of bosonic QAOA we consider the simplest problem possible --- finding the ground state of a single Ising spin in a magnetic field. The cost Hamiltonian for the single Ising spin in a magnetic field is
\begin{align}
    \hat H_C = - K \hat a^{\dagger 2}\hat a^2 + G(\hat a^{\dagger 2} + \hat a^2) + E(\hat a^\dagger + \hat a).
\end{align}
In the simulations we begin from the vacuum and we set $\Delta = K/(\abs{\alpha}^2e^{-2\abs{\alpha}^2})$ and $E=K/(2\alpha)$. The cost Hamiltonian in the computational basis is $\hat H_C=\hat Z$, whose ground state is $\ket{\bar 1}$. From the numerical simulations we obtain a fidelity of $0.52$ for $p=1$ and of $0.785$ for $p=2$. In both cases these results were obtained by evaluation of the expectation value of the cost Hamiltonian on a $(100\times100)^p$-grid.

The low fidelity finds an interpretation in terms of the expectation value landscape for $p=1$, given in \figref[a]{fig:single-ising}. We see that the landscape is very heavily oscillating, hindering optimization. This should also be compared with the same expectation value landscape of QAOA for qubits which appears instead dramatically smoother, see \figref[b]{fig:single-ising}. The fidelity is moreover equal to $1$ for the $p=1$ qubits.
\begin{figure}[t!]
    \centering
    \includegraphics[width=\linewidth]{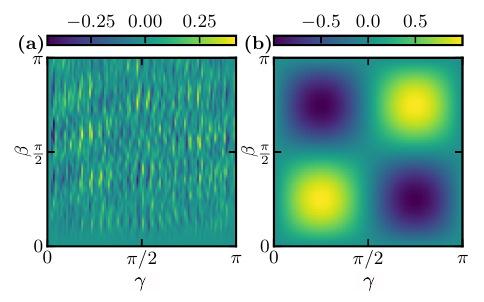}
    \caption{\textbf{(a)} Expectation value landscape of the single-Ising spin for depth $p=1$ in bosonic QAOA. \textbf{(b)} Expectation value landscape of the single-Ising spin for $p=1$ for standard QAOA.}
    \label{fig:single-ising}
\end{figure}

A possible interpretation of this difference in the performance of bosonic QAOA and QAOA resides in the fact that bosonic QAOA starts from the vacuum. Hence, the first iterations of the algorithm are needed just to bring the system onto the qubit computational subspace. In contrast, QAOA implemented with qubits (possibly cat qubits) starts already in the computational subspace. The difficulty of preparing the initial cat state is however somehow hidden in this comparison. Hence in the next subsection we address the preparation of a cat state with bosonic QAOA.
\subsection{\label{app:creating-a-cat}Creating a cat state from vacuum using bosonic QAOA}
\begin{figure}[b!]
    \centering
    \includegraphics{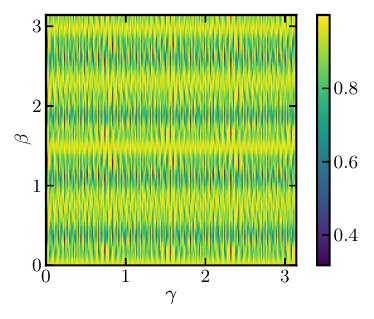}
    \caption{A section of the $p=1$ expectation value landscape for generating a target cat-state with bosonic QAOA. The landscape is highly non-convex.}
    \label{fig:bruteforce}
\end{figure}
Here we investigate the possibility of creating a cat state by starting from the vacuum state and by applying bosonic QAOA. The state evolution is
\begin{align}
    \hat U_p \ket{0}_\mathrm{vac} = \prod_{k=1}^{p}\exp[-i \beta_k\hat H_0] \exp[-i \gamma_k\hat H_1]\ket{0}_\mathrm{vac}
\end{align}
where the two Hamiltonians are given by
\begin{align}
    \hat H_0 = -\Delta \hat a^\dagger \hat a - K \hat a^{\dagger 2}\hat a^2,
\end{align}
and
\begin{align}
    \hat H_1 = - K \hat a^{\dagger 2}\hat a^2 + G(\hat a^{\dagger 2} + \hat a^2).
\end{align}
In the simulations, we use $\Delta=K/(\abs{\alpha}^2e^{-2\abs{\alpha^2}})$ and $G=4K$, we optimize the angles $(\vec\gamma,\vec\beta)$ numerically with respect to minimizing $F(\alpha,\beta)=1-\abs{\braket{C_\alpha^+}{\psi_1(\alpha,\beta)}}^2$. \figref{fig:bruteforce} shows $F(\alpha,\beta)$, where it can be seen that the landscape is highly non-convex and finding the global minimum is challenging. As a result, we obtain a poor fidelity of the variational state with the target cat state, $\abs{\braket{C_\alpha^+}{\psi_1(\alpha^*,\beta^*)}}^2=0.57$.
%
\bibliographystyle{mybibstyle}
\bibliography{references.bib}
%

\end{document}